\documentclass[11pt,a4paper]{article}
\usepackage{jheppub}

\usepackage{amsmath}
\usepackage[dvipsnames]{xcolor}
\usepackage{graphicx}
\usepackage{hyperref}
\usepackage{mathtools}
\usepackage{comment}
\usepackage{cancel}
\usepackage{physics}
\usepackage{mathtools}
\usepackage{adjustbox}
\usepackage{multirow,array,booktabs}

\hypersetup{
 bookmarksnumbered=true,
 colorlinks=true,
 linkcolor=[rgb]{0.098,0.098,0.439},
 citecolor=[rgb]{0.098,0.098,0.439},
 urlcolor=[rgb]{0.098,0.098,0.439}
  }

\title{Second leptogenesis: Unraveling the baryon-lepton asymmetry discrepancy}

\author[a]{YeolLin ChoeJo,}
\author[a]{Kazuki Enomoto,}
\author[a]{Yechan Kim,}
\author[a]{Hye-Sung Lee}
\affiliation[a]{Department of Physics, Korea Advanced Institute of Science and Technology\\ Daejeon 34141, Korea}
\emailAdd{particlephysics@kaist.ac.kr}
\emailAdd{k\_enomoto@kaist.ac.kr}
\emailAdd{cj7801@kaist.ac.kr}
\emailAdd{hyesung.lee@kaist.ac.kr}

\abstract{We propose a novel scenario to explain the matter-antimatter asymmetry by twofold leptogenesis, wherein heavy Majorana neutrinos exhibit temperature-dependent masses and engage in $CP$-violating decays. 
This scenario envisages two distinct phases of leptogenesis: one occurring above the electroweak scale and the other below it.
The sphaleron process converts the first lepton asymmetry to baryon asymmetry, but not the second one due to its decoupling.
This mechanism potentially explains the significant discrepancy between baryon and lepton asymmetries, as suggested by recent observations of Helium-4. Furthermore, our model implies that the present masses of Majorana neutrinos are lighter than the electroweak scale, offering a tangible avenue for experimental verification in various terrestrial settings.}

\begin{document}
\maketitle
\flushbottom

\section{Introduction}

The universe is baryon asymmetric as suggested by the measured baryon-to-photon ratio~\cite{Fields:2019pfx}
\begin{equation}
\label{eq: EtaB}
\eta_B \equiv \frac{n_B-n_{\bar{B}}}{n_{\gamma}} \simeq (6.14 \pm 0.25) \times 10^{-10},
\end{equation}
where $n_B$, $n_{\bar{B}}$, and $n_\gamma$ are the number density of the baryon, antibaryon, and photon, respectively.  
Considering cosmic inflation, $\eta_B$ must be generated after the inflation rather than being an initial condition,~\cite{Murai:2023ntj}, a process known as the baryogenesis~\cite{Sakharov:1967dj}. It is well-established that the standard model (SM) falls short in explaining this~\cite{Gavela:1993ts, Gavela:1994ds, Gavela:1994dt, Huet:1994jb, Kajantie:1996mn, DOnofrio:2015gop}.

Leptogenesis is a plausible new physics scenario to explain $\eta_B$, where the heavy Majorana neutrino $N$ possessing the Yukawa couplings to the lepton doublet $\ell$ and the Higgs doublet $\Phi$, are added to the SM~\cite{Fukugita:1986hr}. 
At the decoupling of $N$ in the early universe, $CP$-violating decays $N \to \ell \Phi$ and $N \to \bar{\ell} \Phi^\dagger$ produce the lepton number ($L$).
It is converted to the baryon number ($B$) via the sphaleron process~\cite{Manton:1983nd, Klinkhamer:1984di}.
$N$ also facilitates the explanation of tiny neutrino masses through the type-I seesaw mechanism~\cite{Minkowski:1977sc, Yanagida:1979as, Gell-Mann:1979vob, Mohapatra:1979ia, Schechter:1980gr}. 

Recently, the EMPRESS experiment reported a new result of the ${}^4\mathrm{He}$ abundance observation~\cite{Matsumoto:2022tlr}. It suggests the large degeneracy parameter of the electron neutrino; $\xi_e = 0.05^{+0.03}_{-0.02}$. 
This implies substantial lepton-to-photon ratio given by~\cite{Kohri:1996ke} 
\begin{equation}
\eta_L \equiv \frac{n_L - n_{\bar L}}{n_\gamma}
\simeq  \sum_l \frac{ g_l \pi^2 }{ 12 \zeta(3) } \left( \frac{ T_l }{ T_\gamma } \right)^3 \xi_l , 
\end{equation}
where $l$ represents all leptons. $g_l$, $\xi_l$, and $T_l$ are the degree of freedom, the degeneracy parameter, and the temperature of $l$, respectively. 
The EMPRESS result suggests
\begin{equation}
\label{eq: EtaL}
\eta_L \simeq \frac{ \pi^2 \sum_{i=e,\mu,\tau} \xi_{\nu_i} }{ 6 \zeta(3) } \left( \frac{ T_\nu }{ T_\gamma } \right)^3 \simeq (7.5^{+4.5}_{-3.0}) \times 10^{-2} , 
\end{equation}
where we have used $(T_\nu/T_\gamma)^3 = 4/11$ and the flavor universality ($\xi_{\nu_e} = \xi_{\nu_\mu} = \xi_{\nu_\tau}$) due to the neutrino oscillations.
Given the universe's electrical neutrality, we disregard the charged lepton asymmetry of the same magnitude as $\eta_B$, as it does not match the observed lepton asymmetry.
See Refs.~\cite{Kawasaki:2022hvx, Burns:2022hkq, Borah:2022uos, Escudero:2022okz, Takahashi:2022cpn, Kasuya:2022cko, Domcke:2022uue, Deng:2023twb, Gao:2023djs} for further discussions on the lepton asymmetry.
While the deviation is at a $2.5\sigma$ level and may not appear excessively large, the suggested value has a significant phenomenological impact.

The lepton asymmetry in Eq.~(\ref{eq: EtaL}) is much larger than the baryon asymmetry in Eq.~(\ref{eq: EtaB}); $\eta_L/\eta_B \simeq 10^8$.
If the sphaleron process occurs frequently, the baryon and lepton asymmetries are made to have the same size, which is proportional to the initial $B-L$~\cite{Khlebnikov:1988sr, Harvey:1990qw}. 
Thus, the traditional baryogenesis scenario cannot explain this large discrepancy. 
The discrepancy seems to suggest that two asymmetries were individually generated by different new physics at different epochs in the early universe.

\begin{figure}[tb]
    \centering
    \includegraphics[width=0.7\textwidth]{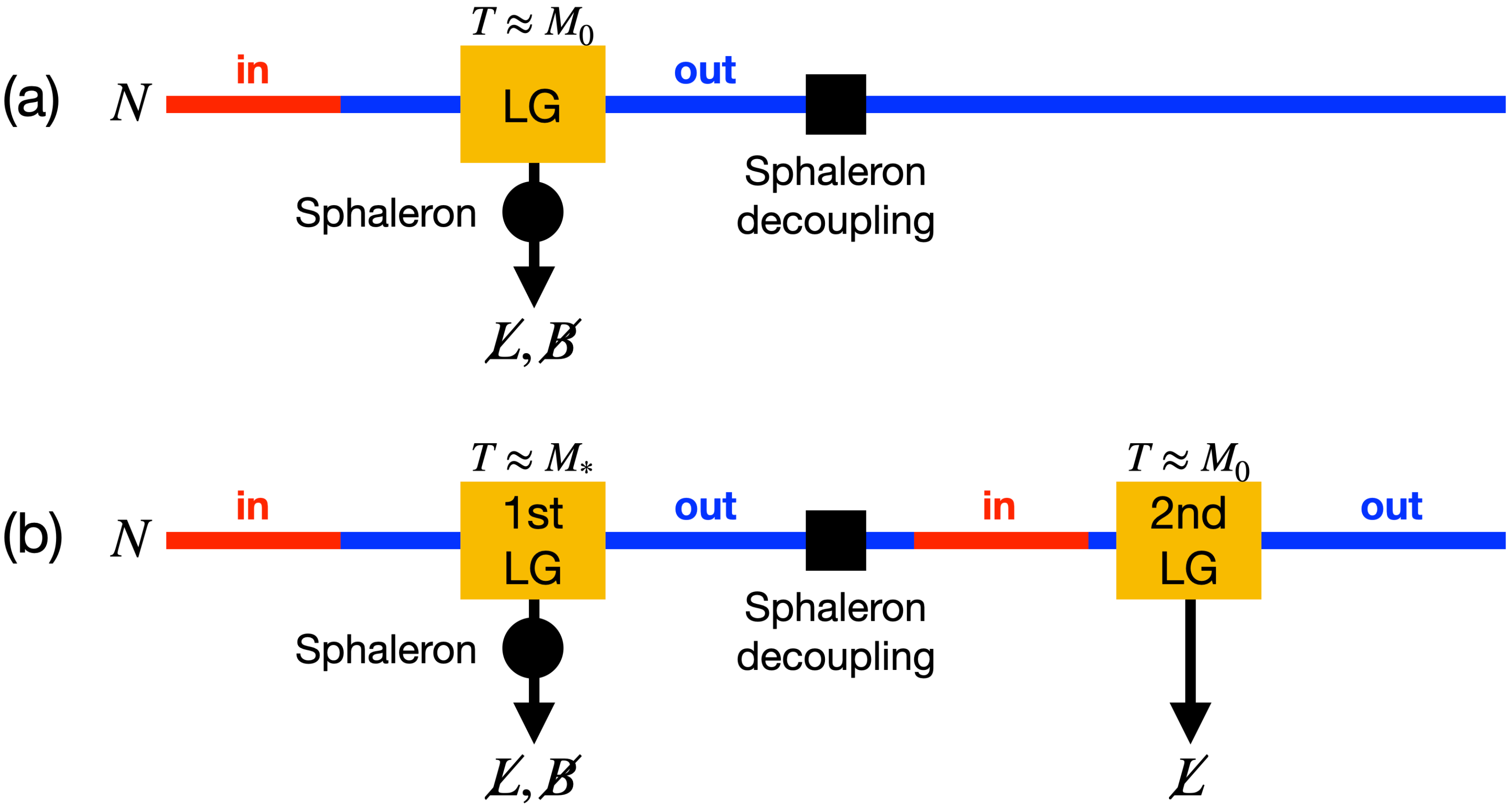}
    \caption{(a) The standard leptogenesis with the constant mass $M_0$.
    In the red (blue) region, $N$ is in (out of) thermal equilibrium. Leptogenesis occurs in the blue region. 
    (b) Leptogenesis may occur twice for the temperature-dependent mass, resulting in a larger lepton asymmetry than the baryon asymmetry. 
    }
    \label{fig: 2ndLG}
\end{figure}

In this paper, we propose a novel, yet simple leptogenesis scenario to explain this discrepancy. It posits two occurrences of leptogenesis in the early universe, driven by the temperature-dependent mass of $N$.
The first leptogenesis takes place prior to the sphaleron decoupling at the electroweak symmetry breaking ($T_{\text{sph}}\simeq 100~\mathrm{GeV}$)~\cite{Kuzmin:1985mm}, and the generated lepton asymmetry is converted to the baryon asymmetry. On the other hand, the second leptogenesis happens below $T_\mathrm{sph}$, allowing the resultant lepton asymmetry to persist into the present universe.

\section{Second leptogenesis}
\label{sec: second leptogenesis}

First, we describe the scenario of the twofold leptogenesis and how it can explain the suggested large discrepancy between the baryon and lepton asymmetries.
This concept is visualized in Figure~\ref{fig: 2ndLG}, alongside a comparison with the standard leptogenesis scenario.

We consider the following case. In the early universe, the heavy neutrino $N$ acquires the temperature-dependent Majorana mass by the new physics effect in addition to the bare Majorana mass $M_0$. The sums of them are denoted by $M(T)$. We assume that $M_0$ is smaller than the electroweak scale; $M_0 < 100~\mathrm{GeV}$.

$M(T)$ behaves as a constant $M_\ast$ $(\gg M_0)$ above the temperature $T_\ast$. 
At $T < T_\ast$, it decreases as a function of temperature.  
When the temperature has dropped enough below a certain temperature $T_N$ $(\ll T_\ast)$, the new physics effect becomes negligibly small, and $M(T) \simeq M_0$ until the current universe. 
A specific new physics to realize this scenario will be discussed later in this paper. 

At high temperatures $T > T_\ast$, $N$ behaves as Majorana fermion with the mass $M_\ast$ and is thermalized via the Yukawa interaction. 
At $T \simeq M_\ast$, the production rate is exponentially suppressed, and $N$ begins to be decoupled. 
The first leptogenesis occurs at this stage. $L$ is generated, and it is converted to $B$, which remains as the baryon asymmetry until the current universe. 

At $T < T_\ast$, $M(T)$ begins to decrease as cooling of the universe; $M(T) \propto T^n$. 
If this decrease is faster than the temperature decrease ($n>1$), 
$M(T)$ can be lower than $T$ at some point, and $N$ can be thermalized again.

As the universe cools further, $M(T)$ behaves as a constant again but much smaller than the mass at the high temperature; $M(T) \simeq M_0 \ll M_\ast$. 
At $T \simeq M_0$, the heavy neutrino is decoupled again. 
The second leptogenesis occurs at this stage, generating the extra lepton number $\Delta L$. 
Since $M_0$ is lower than the electroweak scale, the sphaleron process has already decoupled, and $\Delta L$ remains as the additional lepton asymmetry until the current universe. 

Since the baryon asymmetry is set by $L$, the size of $\Delta L$ has to be much larger than $L$ to explain the large baryon-lepton asymmetry discrepancy.
How can a large enhancement of $\Delta L$ be made at the second leptogenesis? 
It can be realized in a natural way. 
The size of the Yukawa coupling $y$ is proportional to $\sqrt{M_0}$ to reproduce the neutrino mass matrix by the type-I seesaw mechanism~\cite{Casas:2001sr}. 
Thus, the ratio of the production rate $\Gamma_\mathrm{prod} \propto y^2 M(T)$ and the Hubble parameter $H \propto T^2$ is given by $\Gamma_\mathrm{prod} /H \propto M(T) M_0 / T^2$. 
At the first leptogenesis ($T \simeq M_\ast$), the production is much suppressed because $\Gamma_\mathrm{prod}/H \propto M_0/M_\ast \ll 1$. Such a case is referred to as the weak washout~\cite{Davidson:2008bu}, and the generated lepton asymmetry is also suppressed. 
On the other hand, at the second leptogenesis ($T \simeq M_0$), 
the production rate is not suppressed because $\Gamma_\mathrm{prod} /H \propto M(T) M_0 / T^2 |_{T=M_0} \simeq 1$. 
This is the strong washout~\cite{Davidson:2008bu}, and much larger lepton asymmetry can be generated via the second leptogenesis compared to the first one. 
It can naturally explain the large difference between the baryon and lepton asymmetries.

\section{Realization in a wave dark matter model}

Here, we discuss a specific realization of the temperature-dependent mass of $N$.
The neutrino mass variation over cosmic time has been studied a lot in the context of the mass-varying neutrinos in the quintessence dark energy field, where the neutrinos may get their masses from the quintessence field~\cite{Fardon:2003eh, Brookfield:2005bz}.
Lately, there have been many new studies of the mass-varying neutrinos~\cite{Reynoso:2016hjr, Berlin:2016woy, Zhao:2017wmo, Krnjaic:2017zlz, Brdar:2017kbt, Davoudiasl:2018hjw, Liao:2018byh, Capozzi:2018bps, Huang:2018cwo, Farzan:2019yvo, Cline:2019seo, Dev:2020kgz, Losada:2021bxx, Huang:2021kam, Chun:2021ief, Dev:2022bae, Huang:2022wmz, Losada:2022uvr, Brzeminski:2022rkf, Alonso-Alvarez:2023tii, Losada:2023zap, Davoudiasl:2023uiq, Gherghetta:2023myo, ChoeJo:2023ffp, Chen:2023vkq}, including those taking the scalar wave dark matter (DM)~\cite{Hui:2021tkt} in place of the quintessence scalar field.
In this paper, we focus on the case that $M(T)$ is given by the coupling to the scalar wave DM.

The scalar wave DM $\phi$ obeys the following equation of motion with the assumption of spatial homogeneity; 
\begin{equation}
\ddot{\phi} + 3H\dot{\phi} + m_{\phi}^2 \phi = 0,
\end{equation}
where $m_\phi$ is the mass of $\phi$ and is constrained to $3\times10^{-21}$ eV $< m_{\phi} < $ 30 eV~\cite{Hui:2021tkt}. 
The lower bound arises from Lyman-$\alpha$ forest data, and the upper from the de Broglie wavelength exceeding inter-particle separation.

At high temperatures, $\phi$ is fixed at the nonzero initial value due to the large Hubble friction.
Since the $H$ decreases over time, it becomes comparable to $m_\phi$ at the temperature $T_\ast$; $H(T_\ast) = m_\phi$. 
We assume the universe is radiation-dominated in the following. It leads to $T_\ast \simeq \left(m_\phi M_\mathrm{Pl} \sqrt{90/(8\pi^3 g_\ast)} \right)^{1/2}$, where $M_\mathrm{Pl}$ is the Planck mass and $g_\ast = 106.75$ is the effective degree of freedom of the energy density~\cite{Zhao:2017wmo, Brdar:2017kbt, Dev:2022bae, ChoeJo:2023ffp}.

At $T < T_\ast$, $\phi$ coherently oscillates by the mass term; 
\begin{equation}
\label{scalar}
\phi(t) = \frac{\sqrt{2 \rho(t)}}{m_{\phi}} \cos (m_{\phi} t),
\end{equation}
where $\rho(t) = \frac{ 1 }{ 2 } \dot{\phi}^2 + \frac{ 1 }{ 2 } m_\phi^2 \phi^2$ is the energy density of $\phi$.
Since $\rho$ behaves as the matter-like, $\rho \propto a^{-3}$ where $a$ is the scale factor, the oscillation amplitude becomes smaller proportional to $a^{-3/2} \propto T^{3/2}$ as the temperature decreases~\cite{Preskill:1982cy, Abbott:1982af, Dine:1982ah}.

At the current temperature $T_0 \simeq 2.73~\mathrm{K}$~\cite{ParticleDataGroup:2022pth}, the oscillation energy density $\rho_0$ contributes to $\rho_\mathrm{DM}$, the relic energy density of the DM. 
In this paper, we assume that the oscillating $\phi$ contributes to the entire DM. 
It requires the current oscillation amplitude $\phi_0$ to be $\phi_0 = \sqrt{2 \rho_\mathrm{DM}}/m_\phi$. 

The interaction between particles and the wave DM provides the time-dependent mass of the particles.
Some references have investigated leptogenesis with the varying neutrino mass by using quintessence dark energy~\cite{Bi:2003yr, Gu:2004xx}, the neutrino itself as dark energy~\cite{Hati:2015hvq}, or other new physics \cite{Bae:2016zym}. 
Nevertheless, none of these works discussed the possibility of the second leptogenesis.

We assume three Majorana neutrinos $N_i$ ($i=1,2,3$) which couple to the scalar wave DM~\cite{Krnjaic:2017zlz, Dev:2022bae, ChoeJo:2023ffp}. 
The relevant part of the Lagrangian is given by
\begin{equation}
\label{interaction}
    \mathcal{L} = - \frac{1}{2} (M_{0i} + g_i \phi) \overline{N}_i^c N_i + \mathrm{h.c.}
\end{equation}
where $M_{0i}$ and $g_i$ are the bare Majorana masses and coupling constants of $N_i$, respectively.
We do not consider off-diagonal couplings for simplicity.
The cosmic scaling of the Majorana neutrino mass in the wave DM setup was also studied in Refs.~\cite{Krnjaic:2017zlz, Dev:2022bae, ChoeJo:2023ffp}.

The second term in Eq.~(\ref{interaction}) generates the time-dependent mass $M_i(t) = M_{0i} + g_i \phi(t)$. If the oscillation period is much shorter than the time scale of the relevant physics (in our case, $N_i$ decay time), the oscillating term $\phi(t)$ can be approximated by its time average. As a result, we obtain the temperature-dependent mass of $N_i$ as follows;
\begin{equation}
\label{Majorana}
M_i(T) = 
\begin{cases}
M_{0i} + \frac{g_i \phi_0}{\sqrt{2}} \left(\frac{T_\ast}{T_0}\right)^{3/2} & T>T_\ast,    
\\
M_{0i} + \frac{g_i \phi_0}{\sqrt{2}} \left(\frac{T}{T_0}\right)^{3/2} & T_\ast>T.
\end{cases}
\end{equation}

In order to have the twofold leptogenesis scenario described earlier, we consider the case that the first term of $M_i(T)$ is dominant at low temperatures ($T\ll T_\ast$); on the other hand, the second one is at high temperatures ($T\simeq T_\ast$). Then, we can find the temperature $T_{N_i}$ ($\ll T_\ast$), below which the effect of the wave DM (the second term) is negligible compared with the bare mass (the first term). It is evaluated by $M_{0i} = g_i\phi_0(T_{N_i}/T_0)^{3/2}/\sqrt{2}$.

\begin{figure}[t]
    \centering
    \includegraphics[width=0.7\textwidth]{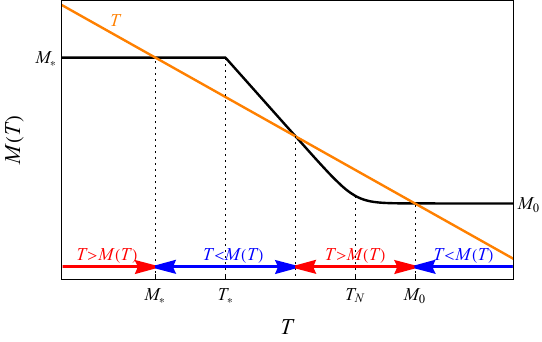}
    \caption{
    The temperature-dependent mass $M(T)$ in Eq.~(\ref{Mcase}) using the time-average approximation (black) in comparison with temperature $T$ (orange). 
    Leptogenesis occurs when the heavy neutrinos decouple (two blue periods). Indices $i$ are suppressed here.
    }
    \label{fig: TempMass}
\end{figure}

\begin{table}[tb]
    \centering
    \begin{tabular}{@{}l*{7}{l}@{}} 
    \toprule
    $m_{\phi}$ & Wave dark matter mass & 
    \\
    \hline
    $M_{\ast i}$ & Majorana mass at $T_*$ &  \\
    $M_{0 i}$ & Bare Majorana mass & \\
    \hline
    $T_*$ & Temperature when oscillation starts &  \\
    $T_{N_i}$ & Temperature when $g_i \phi$ term becomes negligible \\
    $T_0$ & Temperature of the current universe &  \\
    \bottomrule
    \end{tabular}
    \caption{The notations of masses and temperatures.
    The free parameters are only $m_{\phi}, \; M_{\ast i}$ and $M_{0i}$, and others are determined by these parameters.}
    \label{tableMT}
\end{table}

Consequently, the behavior of $M_i(T)$ is described by 
\begin{equation}
\label{Mcase}
M_i(T) \simeq
\begin{cases}
M_{\ast i}  & T > T_\ast, \\
M_{0i} + g_i \frac{ \phi_0 }{\sqrt{2} } \left(\frac{T}{T_0} \right)^{3/2} & T_\ast > T > T_{N_i}, \\
M_{0i} & T_{N_i} > T, 
\end{cases}
\end{equation}
where $M_{\ast i} \equiv M_{0i} + g_i \phi_0(T_\ast/T_0)^{3/2}/\sqrt{2}$.
In Figure~\ref{fig: TempMass}, we illustrate how $M_i(T)$ varies with temperature in the second leptogenesis scenario.
The masses and temperatures are summarized in Table~\ref{tableMT}.

\begin{figure}[tb]
\centering
\includegraphics[width=0.55\textwidth]{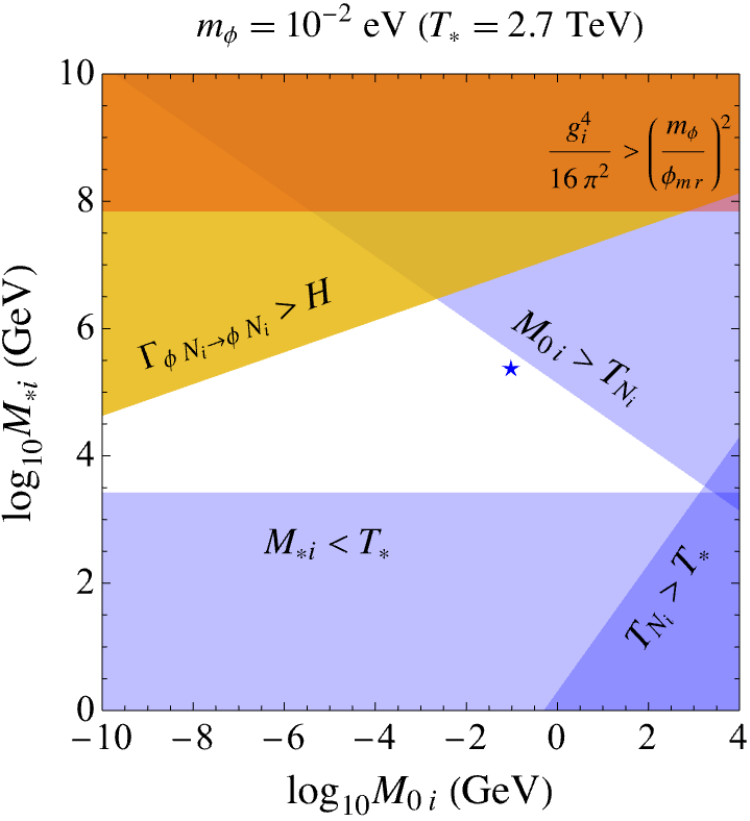}
\caption{The allowed region for the second leptogenesis (blank region) for $m_{\phi} = 10^{-2}$ eV. In the blue regions, the necessary conditions for the second leptogenesis are not satisfied. The red and yellow regions are excluded by theoretical requirements to avoid a dominant quartic coupling and thermalization of $\phi$, respectively. The blue star is the benchmark point for Figure~\ref{fig: NPlot5}.} 
\label{constraints_fig}
\end{figure}

\section{Parameter regions for the second leptogenesis}

Here, we consider constraints on the model parameters. 
The Lagrangian of the model includes three kinds of new parameters $m_\phi$, $M_{0i}$, and $g_i$. We can also choose a more convenient set of the parameters: $T_\ast$, $M_{0i}$, and $M_{\ast i}$. 
Using these parameters, $T_{N_i}$ is evaluated by 
\begin{align}
T_{N_i} = T_\ast \left( \frac{ M_{0i} }{ M_{\ast i} - M_{0i} } \right)^{2/3} \simeq T_\ast \left( \frac{ M_{0i} }{ M_{\ast i} } \right)^{2/3}.
\label{TN}
\end{align}

The necessary conditions for the second leptogenesis are as follows. For the lightest $N_i$,
(i) $M_{\ast i} > T_\ast$: The first decoupling happens earlier than $T_*$. (ii) $T_{N_i} > M_{0i}$: The second decoupling occurs later than $T_{N_i}$. (iii) $T_\ast > T_{N_i}$: The time interval for the temperature-dependent mass should exist, which is equivalent to $M_{\ast i} > 2 M_{0i}$ by Eq.~\eqref{TN}.

Next, we consider theoretical constraints. 
The quartic coupling of $\phi$ is radiatively induced by $g_i$, and it has to be smaller than the mass term at least at the matter-radiation equality $T_{mr} \simeq 1~\mathrm{keV}$, otherwise $\rho(t)$ behaves as $a^{-4}$ not $a^{-3}$~\cite{Dev:2022bae, Davoudiasl:2023uiq, Turner:1983he}. This requires $m_\phi^2/\phi_{mr}^2  > g_i^4/(16\pi^2)$, where $\phi_{mr}$ is the oscillation amplitude at $T_{mr}$. 
In addition, in order to avoid thermalization of $\phi$, the scattering rate has to be smaller than $H(T)$~\cite{Dev:2022bae}. 
We consider two scatterings $\phi \nu \to \phi \nu$ and $\phi N_i \to \phi N_i$. 
The former gives a weaker constraint because $\nu$ couples to $\phi$ only via tiny mixing. The scattering rate of the latter is roughly given by $\Gamma_{\phi N_i \to \phi N_i} \sim g_i^4 T$ when $N_i$ is relativistic. 
Thus, we obtain $g_i^4 > \sqrt{8\pi^3g_\ast/90} \; T/M_\mathrm{Pl}$.

The coupling $g_i$ is also subject to constraints from various experimental studies, such as the Majoron emitting decay~\cite{Doi:1985dx, GERDA:2022ffe}, neutrino free-streaming on the CMB~\cite{Huang:2021kam}, and neutrino oscillations~\cite{Berlin:2016woy} depending on the $m_\phi$ values. 
However, their constraints are weaker than others in the parameter space we are interested in.

In Fig.~\ref{constraints_fig}, we show the allowed parameter regions for the second leptogenesis in the case of $m_\phi = 10^{-2}~\mathrm{eV}$, which corresponds to $T_\ast \simeq 2.7~\mathrm{TeV}$. The blue regions do not satisfy the three conditions required for the second leptogenesis. 
The red and yellow regions are excluded by theoretical constraints to avoid a dominant quartic coupling at $T=T_{mr}$ and to prevent the thermalization of $\phi$, respectively. The experimental constraints are too weak to be shown in the figure. Consequently, the second leptogenesis is expected to occur in the blank regions of the figure. 
In numerical evaluations, we use the average density of the DM $\rho_\mathrm{DM} = 1.2\times 10^{-6}~\mathrm{GeV/cm^3}$, not the local density because we investigate phenomena in the early universe.

The allowed region changes with different values of $m_\phi$. For example, the constraint $\Gamma_{\phi N_i \rightarrow \phi N_i} > H$ becomes stronger for larger $m_\phi$, while the quartic coupling constraint becomes more stringent for smaller $m_\phi$.
Since we consider the scenario where the oscillation of $\phi$ begins before the sphaleron decoupling as explained in Sec.~\ref{sec: second leptogenesis}, $m_\phi$ needs to be larger than $10^{-5}~\mathrm{eV}$, which is derived from $T_\ast > T_\mathrm{sph} \simeq 100~\mathrm{GeV}$. Thus, our relevant mass region is $10^{-5}~\mathrm{eV} < m_\phi < 30~\mathrm{eV}$. We have checked that the allowed region does not change significantly, and we can find a lot of parameter points to achieve the second leptogenesis in this mass region.

\section{Quantitative result of the second leptogenesis}

The asymmetry production in leptogenesis is evaluated by the density matrix equation including the flavor effect~\cite{DeSimone:2006nrs, Blanchet:2006ch, Blanchet:2011xq}. 
Here, we employ the formalism given in Refs.~\cite{Moffat:2018wke, Granelli:2021fyc, Granelli:2023egb} with the addition of the temperature dependence in the masses. 
The equation is given by
\begin{align}
\frac{ \mathrm{d} N_{N_i} }{ \mathrm{d} z } = & - D_i (N_{N_i} - N_{N_i}^\mathrm{eq} ),
\label{Neqnres}
\\
\frac{ \mathrm{d} N_{\alpha \beta} }{ \mathrm{d} z } = & \sum_i \left[ \varepsilon^{(i)}_{\alpha \beta} D_i (N_{N_i} - N_{N_i}^\mathrm{eq} )
- \frac{1}{2} W_i \{ P_i, N \}_{\alpha \beta} \right] \nonumber \\
& - \frac{\Gamma_{\tau}}{Hz} [I_{\tau}, [I_{\tau}, N]]_{\alpha\beta}
- \frac{\Gamma_{\mu}}{Hz} [I_{\mu}, [I_{\mu}, N]]_{\alpha\beta},
\label{NBLeqnres}
\end{align}
where $z=M_{01}/T$, $i=1$, 2, 3, and $\alpha$, $\beta$ = $e$, $\mu$, $\tau$. 
$N_{N_i}$ and the diagonal terms $N_{\alpha \alpha}$ are the number of $N_i$ and $B/3 - L_\alpha$, where $L_\alpha$ is the lepton number for each flavor, respectively, in a portion of the comoving volume that contains one photon at the era when $N_i$ is relativistic and in thermal equilibrium~\cite{Buchmuller:2002rq, Buchmuller:2004nz}. 
The off-diagonal terms $N_{\alpha \beta}$ ($\alpha \neq \beta$) represent the coherence between the flavors. 
The number of the total $B-L$ is given by $N_{B-L} = \sum_{\alpha } N_{\alpha \alpha}$. 
The term $D_i$ accounts for the decay and inverse decay of $N_i$. 
The washout effect in $B-L$ asymmetry is described by $W_i$. 
We consider the washout effect from the inverse decay and neglect one from other lepton-number-violating processes. 
$D_i$ and $W_i$ are given by~\cite{Buchmuller:2004nz} 
\begin{align}
& D_i = \frac{ (yy^\dagger)_{ii} }{ 8 \pi H z } M_i(T) \frac{ K_1 \bigl(M_i(T)/T \bigr) }{ K_2 \bigl( M_i(T)/T \bigr) }, \\
& W_i = \frac{ 2 }{ 3 } D_i N_{N_i}^\mathrm{eq},
\end{align}
where $N_{N_i}^\mathrm{eq} = (3/8)\bigl(M_i(T)/T\bigr)^2 K_2\bigl(M_i(T)/T\bigr)$ is the equilibrium value of $N_{N_i}$, and $y$ is the Yukawa matrix for the interaction among the lepton doublet, the Higgs doublet, and the heavy neutrinos.
$K_n$ is a modified $n$-th Bessel function of the second kind.
We note that the temperature dependence of the mass is included in these terms. 
$P_i$ is the projection matrix constructed with the Yukawa matrix.
Decoherence effects by the interchanges between the left-handed and right-handed leptons are described by the double commutator terms, where $\Gamma_{\mu(\tau)} \simeq 8 \times 10^{-3} (\sqrt{2}m_{\mu(\tau)}/v)^2 T$~\cite{Granelli:2021fyc} is the rate of the process involving $\mu$($\tau$), $I_{\mu}= \text{diag} (0,1,0)$, and $I_{\tau} = \text{diag}(0,0,1)$.

Since $M_{0i}$ in our scenario is much lighter than the Davidson-Ibarra bound, $M_{0i} \gtrsim 10^8~\mathrm{GeV}$~\cite{Davidson:2002qv}, for the heavy neutrinos with hierarchical masses, we consider resonant leptogenesis, where the heavy neutrinos possess very close masses~\cite{Pilaftsis:1998pd, Pilaftsis:2003gt}. 
The $CP$ asymmetry parameter $\varepsilon^{(i)}_{\alpha \beta}$ is divided into contributions from vertex and self-energy diagrams~\cite{Davidson:2008bu}. The self-energy part $\varepsilon^{S(i)}_{\alpha \beta}$ is resonantly enhanced with degenerate masses and can dominate the $CP$ asymmetry parameter. $\varepsilon^{S(i)}_{\alpha \beta}$ is given by
\begin{align}
\label{epsilon}
\varepsilon^{S(i)}_{\alpha \beta} =& \frac{1}{16\pi   (yy^{\dagger})_{ii}} 
\sum_{j\neq i} \Bigl\{ 
	 i \bigl[ y^\ast_{i\alpha} y_{j \beta} (yy^{\dagger})_{ji} - y^{}_{i \beta} y^\ast_{j \alpha} (yy^{\dagger})_{ij} \bigr]
 \frac{M_j}{M_i} \nonumber \\
 	& + i  \bigl[ y^\ast_{i\alpha} y_{j \beta} (yy^{\dagger})_{ij} - y^{}_{i \beta} y^\ast_{j \alpha} (yy^{\dagger})_{ji} \bigr] 	 
\Bigr\} \frac{ (M_j^2 - M_i^2)M_i^2 }{ (M_j^2 - M_i^2)^2 + M_i^4 \Gamma_j^2 / M_j^2}, 
\end{align}
where $M_j$ and $\Gamma_j$ are the temperature-dependent mass and decay rate of $N_j$, respectively.
Thus, $\varepsilon^{(i)}_{\alpha \beta}$ is significantly enhanced when the resonant condition, $|M_j - M_i| \simeq \Gamma_j/2$, is satisfied.

As a benchmark point to evaluate $N_{N_1}$ and $N_{B-L}$, 
we assume the following input parameters: $m_{\phi} = 10^{-2}$ eV ($T_{\ast} \simeq 2.7$ TeV), $M_{01} = 0.1$ GeV, and $M_{\ast 1} = 2.4 \times 10^5$ GeV.
The current masses of $N_2$ and $N_3$ are chosen to satisfy the resonant condition at the second leptogenesis, $\Delta M_{12} \equiv M_{02} - M_{01} = 0.5 \times 10^{-19}$ GeV and $\Delta M_{13} \equiv M_{03} - M_{01} = 4.0 \times 10^{-19}$ GeV.
$M_{\ast i}$ ($i=2,3$) are determined by imposing $M_{\ast i}/M_{\ast 1} = M_{0i}/M_{01}$ by which the resonant condition is also satisfied at the first leptogenesis.

The Yukawa matrix $y$ is set to satisfy the neutrino oscillation data with the normal ordering masses~\cite{ParticleDataGroup:2022pth} by using the Casas-Ibarra parametrization \cite{Casas:2001sr}, $y = \sqrt{2} \hat{M}_N^{1/2} R \hat{m}_{\nu}^{1/2} U^{\dagger} / v$, where $\hat{M}_N$ and $\hat{m}_{\nu}$ are the diagonal mass matrices of the heavy and active neutrinos, respectively, $U$ is the PMNS matrix~\cite{Pontecorvo:1957cp, Maki:1962mu}, $v$ is the vacuum expectation value of the Higgs, and $R$ is a complex orthogonal matrix. 
We note that $\hat{M}_N$ is evaluated at the current temperature. 
Six parameters in $y$ are undetermined by the neutrino oscillation data: the lightest neutrino mass $m_{\nu_1}$, two Majorana phases $\alpha_1, \alpha_2$ in the PMNS matrix $U$, and three complex phases $\omega_1$, $\omega_2$, $\omega_3$ in $R$.
The notations of the Majorana phases, and the complex phases follow Refs.~\cite{ParticleDataGroup:2022pth} and \cite{Moffat:2018wke}, respectively. We assume the following values for them; $m_{\nu_1} = 0~\mathrm{eV}$, $\alpha_1 = \alpha_2 = 0$, $\omega_1 = \omega_2 = 0$, and $\omega_3 = 0.2e^{-i\pi/4}$.

\begin{figure}[tb]
    \centering
    \includegraphics[width=0.7\textwidth]{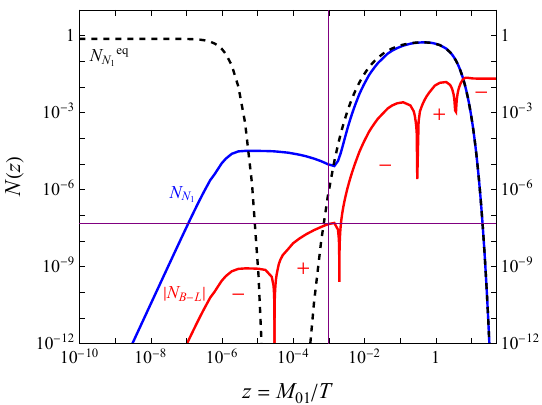}
    \caption{
    $N_{N_1}$ and $|N_{B-L}|$ for resonant leptogenesis with the temperature-dependent masses of the neutrinos on the benchmark point described in the text.
    Here, $|N_{B-L}|$ is the sum of the $B-L$ asymmetry deposited in each flavor, and its sign is represented together.
    The baryon asymmetry is determined after the sphaleron decoupling ($T \simeq 100~\mathrm{GeV}$ corresponding to $z \simeq 10^{-3}$ with $M_{01} = 0.1~\mathrm{GeV}$) with positive $N_{B-L}$, but the lepton asymmetry is fixed only after the second leptogenesis ($z \simeq 1$) with negative $N_{B-L}$.
    The baryogenesis fitting at the sphaleron decoupling ($z \simeq 10^{-3}$) is shown together (thin lines).
    }
    \label{fig: NPlot5}
\end{figure}

We consider thermal leptogenesis with the initial condition $N_{N_1} = N_{B-L} = 0$ at $z = 10^{-10}$ ($T = 10^9~\mathrm{GeV}$). 
Figure~\ref{fig: NPlot5} shows the behavior of $N_{N_1}$ and $N_{B-L}$ with the above inputs. 
The first leptogenesis occurs a little later than $T = M_{\ast 1}$ ($z \simeq 10^6$) because of weak washout~\cite{Buchmuller:2004nz}. 
The produced lepton number is converted to the baryon number by the sphaleron process, and it is fixed after the sphaleron decoupling. 
For simplicity, we assume that the decoupling occurs instantaneously, and the final baryon number is determined by $N_{B-L}$ at the temperature $T = 100~\mathrm{GeV}$ ($z = 10^3$) shown by the purple lines in Figure~\ref{fig: NPlot5}. 

The baryon-to-photon ratio can be obtained by $\eta_B = a_{\text{sph}} N_{B-L} /f$ at the sphaleron decoupling, where $a_{\text{sph}}$ and $f$ are the sphaleron conversion rate~\cite{Harvey:1990qw, Khlebnikov:1988sr} and the photon dilution factor~\cite{Buchmuller:2002rq} due to the increase of the photon number by annihilation of particles from the first leptogenesis till the BBN, respectively. We use $a_{\text{sph}} = 28/79$, the value for the SM plasma. Since the heavy neutrinos have nearly degenerate masses, we evaluate $f$ including the effect of $N_1$, $N_2$, and $N_3$, not only $N_1$. 
Then, we have $f = 1232/43$.\footnote{With $m$ generations of the heavy neutrinos, $f=11(427+7m)/172$, which leads to the commonly used value $f=2387/86$ in the case of $m=1$~\cite{Buchmuller:2002rq}.}
As a result, we obtain $\eta_B \simeq 6.14 \times 10^{-10}$, which is consistent within $1\sigma$ level with the current observations~\cite{Fields:2019pfx, Planck:2018vyg}.
Also, the positive $N_{B-L}$ at the sphaleron decoupling in our benchmark point provides the correct sign for the baryon asymmetry.

The second leptogenesis commences shortly after $T = M_{01}$ (or $z = 1$), due to the effects of strong washout~\cite{Buchmuller:2004nz}. The lepton number generated during this period persists into the current universe, as all lepton-number-violating processes have ceased. In this scenario, the photon dilution factor $f'$ is calculated based on the change in photon numbers from the second leptogenesis to the era of BBN, rather than from the first leptogenesis.
This is because the $N_1$ species is relativistic and returns to thermal equilibrium before the onset of the second leptogenesis. We adopt $f' = 176/43$, where the thermal bath consists of $e^\pm$, $\nu_\ell$, $\gamma$, and $N_i$ before the second leptogenesis.\footnote{With $m$ generations of the heavy neutrinos, $f^\prime = 11(43+7m)/172$.}
Consequently, we derive a lepton asymmetry value of $ \eta_{L} = N_{B-L}/f' \simeq 5.0 \times 10^{-3}$.
The lepton asymmetry is flavor-universal due to neutrino oscillation, so we consider the summation of the $B-L$ asymmetry across all flavors, rather than focusing solely on the electron component. Since $N_{B-L}$ is negative in the late epoch of our benchmark model, a positive lepton asymmetry is also guaranteed.

This demonstrates that the second leptogenesis can significantly amplify lepton asymmetry, increasing it by several orders of magnitude from the baryon asymmetry. In our analysis, the benchmark point, which was not fully optimized, already indicates $\eta_L \sim 10^{-3}$. This is remarkably close to the EMPRESS data, which suggests $\eta_L \sim 10^{-2}$, and represents a substantial deviation from $\eta_B \sim 10^{-10}$.
A more refined analysis or the addition of more Majorana neutrinos might aid in reconciling the slight discrepancy from the observed values.

\section{Summary and outlook}
It is notable that a significant deviation of the lepton asymmetry from the baryon asymmetry can be explained in a rather simple framework of the second leptogenesis.
This scenario allows only larger lepton asymmetry than the baryon asymmetry, not the other way around, in accordance with the measurement.
A more comprehensive study will follow in the subsequent work.
In the future, there will be increased CMB data from the Simons Observatory \cite{SimonsObservatory:2018koc} and CMB-S4 \cite{Abazajian:2019eic} that can either confirm or refute the discrepancy \cite{Escudero:2022okz}.

\acknowledgments
This work was partly supported by the National Research Foundation of Korea (Grant No. NRF-2021R1A2C2009718).

\bibliographystyle{JHEP}
\bibliography{ref.bib}

\providecommand{\href}[2]{#2}\begingroup\raggedright\begin{thebibliography}{10}

\bibitem{Fields:2019pfx}
B.D.~Fields, K.A.~Olive, T.-H.~Yeh and C.~Young, \emph{{Big-Bang
  Nucleosynthesis after Planck}},
  \href{https://doi.org/10.1088/1475-7516/2020/03/010}{\emph{JCAP} {\bfseries
  03} (2020) 010} [\href{https://arxiv.org/abs/1912.01132}{{\ttfamily
  1912.01132}}].

\bibitem{Murai:2023ntj}
K.~Murai, F.~Takahashi, M.~Yamada and W.~Yin, \emph{{Can baryon asymmetry be
  explained by a large initial value before inflation?}},
  \href{https://doi.org/10.1103/PhysRevD.108.083518}{\emph{Phys. Rev. D}
  {\bfseries 108} (2023) 083518}
  [\href{https://arxiv.org/abs/2307.03049}{{\ttfamily 2307.03049}}].

\bibitem{Sakharov:1967dj}
A.D.~Sakharov, \emph{{Violation of CP Invariance, C asymmetry, and baryon
  asymmetry of the universe}},
  \href{https://doi.org/10.1070/PU1991v034n05ABEH002497}{\emph{Pisma Zh. Eksp.
  Teor. Fiz.} {\bfseries 5} (1967) 32}.

\bibitem{Gavela:1993ts}
M.B.~Gavela, P.~Hernandez, J.~Orloff and O.~Pene, \emph{{Standard model CP
  violation and baryon asymmetry}},
  \href{https://doi.org/10.1142/S0217732394000629}{\emph{Mod. Phys. Lett. A}
  {\bfseries 9} (1994) 795}
  [\href{https://arxiv.org/abs/hep-ph/9312215}{{\ttfamily hep-ph/9312215}}].

\bibitem{Gavela:1994ds}
M.B.~Gavela, M.~Lozano, J.~Orloff and O.~Pene, \emph{{Standard model CP
  violation and baryon asymmetry. Part 1: Zero temperature}},
  \href{https://doi.org/10.1016/0550-3213(94)00409-9}{\emph{Nucl. Phys. B}
  {\bfseries 430} (1994) 345}
  [\href{https://arxiv.org/abs/hep-ph/9406288}{{\ttfamily hep-ph/9406288}}].

\bibitem{Gavela:1994dt}
M.B.~Gavela, P.~Hernandez, J.~Orloff, O.~Pene and C.~Quimbay, \emph{{Standard
  model CP violation and baryon asymmetry. Part 2: Finite temperature}},
  \href{https://doi.org/10.1016/0550-3213(94)00410-2}{\emph{Nucl. Phys. B}
  {\bfseries 430} (1994) 382}
  [\href{https://arxiv.org/abs/hep-ph/9406289}{{\ttfamily hep-ph/9406289}}].

\bibitem{Huet:1994jb}
P.~Huet and E.~Sather, \emph{{Electroweak baryogenesis and standard model CP
  violation}}, \href{https://doi.org/10.1103/PhysRevD.51.379}{\emph{Phys. Rev.
  D} {\bfseries 51} (1995) 379}
  [\href{https://arxiv.org/abs/hep-ph/9404302}{{\ttfamily hep-ph/9404302}}].

\bibitem{Kajantie:1996mn}
K.~Kajantie, M.~Laine, K.~Rummukainen and M.E.~Shaposhnikov, \emph{{Is there a~
  hot electroweak phase transition at $m_H \gtrsim m_W$?}},
  \href{https://doi.org/10.1103/PhysRevLett.77.2887}{\emph{Phys. Rev. Lett.}
  {\bfseries 77} (1996) 2887}
  [\href{https://arxiv.org/abs/hep-ph/9605288}{{\ttfamily hep-ph/9605288}}].

\bibitem{DOnofrio:2015gop}
M.~D'Onofrio and K.~Rummukainen, \emph{{Standard model cross-over on the
  lattice}}, \href{https://doi.org/10.1103/PhysRevD.93.025003}{\emph{Phys. Rev.
  D} {\bfseries 93} (2016) 025003}
  [\href{https://arxiv.org/abs/1508.07161}{{\ttfamily 1508.07161}}].

\bibitem{Fukugita:1986hr}
M.~Fukugita and T.~Yanagida, \emph{{Baryogenesis Without Grand Unification}},
  \href{https://doi.org/10.1016/0370-2693(86)91126-3}{\emph{Phys. Lett. B}
  {\bfseries 174} (1986) 45}.

\bibitem{Manton:1983nd}
N.S.~Manton, \emph{{Topology in the Weinberg-Salam Theory}},
  \href{https://doi.org/10.1103/PhysRevD.28.2019}{\emph{Phys. Rev. D}
  {\bfseries 28} (1983) 2019}.

\bibitem{Klinkhamer:1984di}
F.R.~Klinkhamer and N.S.~Manton, \emph{{A Saddle Point Solution in the
  Weinberg-Salam Theory}},
  \href{https://doi.org/10.1103/PhysRevD.30.2212}{\emph{Phys. Rev. D}
  {\bfseries 30} (1984) 2212}.

\bibitem{Minkowski:1977sc}
P.~Minkowski, \emph{{$\mu \to e\gamma$ at a Rate of One Out of $10^{9}$ Muon
  Decays?}}, \href{https://doi.org/10.1016/0370-2693(77)90435-X}{\emph{Phys.
  Lett. B} {\bfseries 67} (1977) 421}.

\bibitem{Yanagida:1979as}
T.~Yanagida, \emph{{Horizontal gauge symmetry and masses of neutrinos}},
  {\emph{Conf. Proc. C} {\bfseries 7902131} (1979) 95}.

\bibitem{Gell-Mann:1979vob}
M.~Gell-Mann, P.~Ramond and R.~Slansky, \emph{{Complex Spinors and Unified
  Theories}}, {\emph{Conf. Proc. C} {\bfseries 790927} (1979) 315}
  [\href{https://arxiv.org/abs/1306.4669}{{\ttfamily 1306.4669}}].

\bibitem{Mohapatra:1979ia}
R.N.~Mohapatra and G.~Senjanovic, \emph{{Neutrino Mass and Spontaneous Parity
  Nonconservation}},
  \href{https://doi.org/10.1103/PhysRevLett.44.912}{\emph{Phys. Rev. Lett.}
  {\bfseries 44} (1980) 912}.

\bibitem{Schechter:1980gr}
J.~Schechter and J.W.F.~Valle, \emph{{Neutrino Masses in SU(2) x U(1)
  Theories}}, \href{https://doi.org/10.1103/PhysRevD.22.2227}{\emph{Phys. Rev.
  D} {\bfseries 22} (1980) 2227}.

\bibitem{Matsumoto:2022tlr}
A.~Matsumoto et~al., \emph{{EMPRESS. VIII. A New Determination of Primordial He
  Abundance with Extremely Metal-poor Galaxies: A Suggestion of the Lepton
  Asymmetry and Implications for the Hubble Tension}},
  \href{https://doi.org/10.3847/1538-4357/ac9ea1}{\emph{Astrophys. J.}
  {\bfseries 941} (2022) 167}
  [\href{https://arxiv.org/abs/2203.09617}{{\ttfamily 2203.09617}}].

\bibitem{Kohri:1996ke}
K.~Kohri, M.~Kawasaki and K.~Sato, \emph{{Big bang nucleosynthesis and lepton
  number asymmetry in the universe}},
  \href{https://doi.org/10.1086/512793}{\emph{Astrophys. J.} {\bfseries 490}
  (1997) 72} [\href{https://arxiv.org/abs/astro-ph/9612237}{{\ttfamily
  astro-ph/9612237}}].

\bibitem{Kawasaki:2022hvx}
M.~Kawasaki and K.~Murai, \emph{{Lepton asymmetric universe}},
  \href{https://doi.org/10.1088/1475-7516/2022/08/041}{\emph{JCAP} {\bfseries
  08} (2022) 041} [\href{https://arxiv.org/abs/2203.09713}{{\ttfamily
  2203.09713}}].

\bibitem{Burns:2022hkq}
A.-K.~Burns, T.M.P.~Tait and M.~Valli, \emph{{Indications for a Nonzero Lepton
  Asymmetry from Extremely Metal-Poor Galaxies}},
  \href{https://doi.org/10.1103/PhysRevLett.130.131001}{\emph{Phys. Rev. Lett.}
  {\bfseries 130} (2023) 131001}
  [\href{https://arxiv.org/abs/2206.00693}{{\ttfamily 2206.00693}}].

\bibitem{Borah:2022uos}
D.~Borah and A.~Dasgupta, \emph{{Large neutrino asymmetry from TeV scale
  leptogenesis}},
  \href{https://doi.org/10.1103/PhysRevD.108.035015}{\emph{Phys. Rev. D}
  {\bfseries 108} (2023) 035015}
  [\href{https://arxiv.org/abs/2206.14722}{{\ttfamily 2206.14722}}].

\bibitem{Escudero:2022okz}
M.~Escudero, A.~Ibarra and V.~Maura, \emph{{Primordial lepton asymmetries in
  the precision cosmology era: Current status and future sensitivities from BBN
  and the CMB}}, \href{https://doi.org/10.1103/PhysRevD.107.035024}{\emph{Phys.
  Rev. D} {\bfseries 107} (2023) 035024}
  [\href{https://arxiv.org/abs/2208.03201}{{\ttfamily 2208.03201}}].

\bibitem{Takahashi:2022cpn}
T.~Takahashi and S.~Yamashita, \emph{{Big bang nucleosynthesis and early dark
  energy in light of the EMPRESS Yp results and the H0 tension}},
  \href{https://doi.org/10.1103/PhysRevD.107.103520}{\emph{Phys. Rev. D}
  {\bfseries 107} (2023) 103520}
  [\href{https://arxiv.org/abs/2211.04087}{{\ttfamily 2211.04087}}].

\bibitem{Kasuya:2022cko}
S.~Kasuya, M.~Kawasaki and K.~Murai, \emph{{Enhancement of second-order
  gravitational waves at Q-ball decay}},
  \href{https://doi.org/10.1088/1475-7516/2023/05/053}{\emph{JCAP} {\bfseries
  05} (2023) 053} [\href{https://arxiv.org/abs/2212.13370}{{\ttfamily
  2212.13370}}].

\bibitem{Domcke:2022uue}
V.~Domcke, K.~Kamada, K.~Mukaida, K.~Schmitz and M.~Yamada, \emph{{New
  Constraint on Primordial Lepton Flavor Asymmetries}},
  \href{https://doi.org/10.1103/PhysRevLett.130.261803}{\emph{Phys. Rev. Lett.}
  {\bfseries 130} (2023) 261803}
  [\href{https://arxiv.org/abs/2208.03237}{{\ttfamily 2208.03237}}].

\bibitem{Deng:2023twb}
S.~Deng and L.~Bian, \emph{{Constraints on new physics around the MeV scale
  with cosmological observations}},
  \href{https://doi.org/10.1103/PhysRevD.108.063516}{\emph{Phys. Rev. D}
  {\bfseries 108} (2023) 063516}
  [\href{https://arxiv.org/abs/2304.06576}{{\ttfamily 2304.06576}}].

\bibitem{Gao:2023djs}
F.~Gao, J.~Harz, C.~Hati, Y.~Lu, I.M.~Oldengott and G.~White, \emph{{Sphaleron
  freeze-in baryogenesis with gravitational waves from the QCD transition}},
  \href{https://arxiv.org/abs/2309.00672}{{\ttfamily 2309.00672}}.

\bibitem{Khlebnikov:1988sr}
S.Y.~Khlebnikov and M.E.~Shaposhnikov, \emph{{The Statistical Theory of
  Anomalous Fermion Number Nonconservation}},
  \href{https://doi.org/10.1016/0550-3213(88)90133-2}{\emph{Nucl. Phys. B}
  {\bfseries 308} (1988) 885}.

\bibitem{Harvey:1990qw}
J.A.~Harvey and M.S.~Turner, \emph{{Cosmological baryon and lepton number in
  the presence of electroweak fermion number violation}},
  \href{https://doi.org/10.1103/PhysRevD.42.3344}{\emph{Phys. Rev. D}
  {\bfseries 42} (1990) 3344}.

\bibitem{Kuzmin:1985mm}
V.A.~Kuzmin, V.A.~Rubakov and M.E.~Shaposhnikov, \emph{{On the Anomalous
  Electroweak Baryon Number Nonconservation in the Early Universe}},
  \href{https://doi.org/10.1016/0370-2693(85)91028-7}{\emph{Phys. Lett. B}
  {\bfseries 155} (1985) 36}.

\bibitem{Casas:2001sr}
J.A.~Casas and A.~Ibarra, \emph{{Oscillating neutrinos and $\mu \to e,
  \gamma$}}, \href{https://doi.org/10.1016/S0550-3213(01)00475-8}{\emph{Nucl.
  Phys. B} {\bfseries 618} (2001) 171}
  [\href{https://arxiv.org/abs/hep-ph/0103065}{{\ttfamily hep-ph/0103065}}].

\bibitem{Davidson:2008bu}
S.~Davidson, E.~Nardi and Y.~Nir, \emph{{Leptogenesis}},
  \href{https://doi.org/10.1016/j.physrep.2008.06.002}{\emph{Phys. Rept.}
  {\bfseries 466} (2008) 105}
  [\href{https://arxiv.org/abs/0802.2962}{{\ttfamily 0802.2962}}].

\bibitem{Fardon:2003eh}
R.~Fardon, A.E.~Nelson and N.~Weiner, \emph{{Dark energy from mass varying
  neutrinos}}, \href{https://doi.org/10.1088/1475-7516/2004/10/005}{\emph{JCAP}
  {\bfseries 10} (2004) 005}
  [\href{https://arxiv.org/abs/astro-ph/0309800}{{\ttfamily
  astro-ph/0309800}}].

\bibitem{Brookfield:2005bz}
A.W.~Brookfield, C.~van~de Bruck, D.F.~Mota and D.~Tocchini-Valentini,
  \emph{{Cosmology of mass-varying neutrinos driven by quintessence: theory and
  observations}}, \href{https://doi.org/10.1103/PhysRevD.73.083515}{\emph{Phys.
  Rev. D} {\bfseries 73} (2006) 083515}
  [\href{https://arxiv.org/abs/astro-ph/0512367}{{\ttfamily
  astro-ph/0512367}}].

\bibitem{Reynoso:2016hjr}
M.M.~Reynoso and O.A.~Sampayo, \emph{{Propagation of high-energy neutrinos in a
  background of ultralight scalar dark matter}},
  \href{https://doi.org/10.1016/j.astropartphys.2016.05.004}{\emph{Astropart.
  Phys.} {\bfseries 82} (2016) 10}
  [\href{https://arxiv.org/abs/1605.09671}{{\ttfamily 1605.09671}}].

\bibitem{Berlin:2016woy}
A.~Berlin, \emph{{Neutrino Oscillations as a Probe of Light Scalar Dark
  Matter}}, \href{https://doi.org/10.1103/PhysRevLett.117.231801}{\emph{Phys.
  Rev. Lett.} {\bfseries 117} (2016) 231801}
  [\href{https://arxiv.org/abs/1608.01307}{{\ttfamily 1608.01307}}].

\bibitem{Zhao:2017wmo}
Y.~Zhao, \emph{{Cosmology and time dependent parameters induced by a misaligned
  light scalar}}, \href{https://doi.org/10.1103/PhysRevD.95.115002}{\emph{Phys.
  Rev. D} {\bfseries 95} (2017) 115002}
  [\href{https://arxiv.org/abs/1701.02735}{{\ttfamily 1701.02735}}].

\bibitem{Krnjaic:2017zlz}
G.~Krnjaic, P.A.N.~Machado and L.~Necib, \emph{{Distorted neutrino oscillations
  from time varying cosmic fields}},
  \href{https://doi.org/10.1103/PhysRevD.97.075017}{\emph{Phys. Rev. D}
  {\bfseries 97} (2018) 075017}
  [\href{https://arxiv.org/abs/1705.06740}{{\ttfamily 1705.06740}}].

\bibitem{Brdar:2017kbt}
V.~Brdar, J.~Kopp, J.~Liu, P.~Prass and X.-P.~Wang, \emph{{Fuzzy dark matter
  and nonstandard neutrino interactions}},
  \href{https://doi.org/10.1103/PhysRevD.97.043001}{\emph{Phys. Rev. D}
  {\bfseries 97} (2018) 043001}
  [\href{https://arxiv.org/abs/1705.09455}{{\ttfamily 1705.09455}}].

\bibitem{Davoudiasl:2018hjw}
H.~Davoudiasl, G.~Mohlabeng and M.~Sullivan, \emph{{Galactic Dark Matter
  Population as the Source of Neutrino Masses}},
  \href{https://doi.org/10.1103/PhysRevD.98.021301}{\emph{Phys. Rev. D}
  {\bfseries 98} (2018) 021301}
  [\href{https://arxiv.org/abs/1803.00012}{{\ttfamily 1803.00012}}].

\bibitem{Liao:2018byh}
J.~Liao, D.~Marfatia and K.~Whisnant, \emph{{Light scalar dark matter at
  neutrino oscillation experiments}},
  \href{https://doi.org/10.1007/JHEP04(2018)136}{\emph{JHEP} {\bfseries 04}
  (2018) 136} [\href{https://arxiv.org/abs/1803.01773}{{\ttfamily
  1803.01773}}].

\bibitem{Capozzi:2018bps}
F.~Capozzi, I.M.~Shoemaker and L.~Vecchi, \emph{{Neutrino Oscillations in Dark
  Backgrounds}},
  \href{https://doi.org/10.1088/1475-7516/2018/07/004}{\emph{JCAP} {\bfseries
  07} (2018) 004} [\href{https://arxiv.org/abs/1804.05117}{{\ttfamily
  1804.05117}}].

\bibitem{Huang:2018cwo}
G.-Y.~Huang and N.~Nath, \emph{{Neutrinophilic Axion-Like Dark Matter}},
  \href{https://doi.org/10.1140/epjc/s10052-018-6391-y}{\emph{Eur. Phys. J. C}
  {\bfseries 78} (2018) 922}
  [\href{https://arxiv.org/abs/1809.01111}{{\ttfamily 1809.01111}}].

\bibitem{Farzan:2019yvo}
Y.~Farzan, \emph{{Ultra-light scalar saving the 3 + 1 neutrino scheme from the
  cosmological bounds}},
  \href{https://doi.org/10.1016/j.physletb.2019.134911}{\emph{Phys. Lett. B}
  {\bfseries 797} (2019) 134911}
  [\href{https://arxiv.org/abs/1907.04271}{{\ttfamily 1907.04271}}].

\bibitem{Cline:2019seo}
J.M.~Cline, \emph{{Viable secret neutrino interactions with ultralight dark
  matter}}, \href{https://doi.org/10.1016/j.physletb.2019.135182}{\emph{Phys.
  Lett. B} {\bfseries 802} (2020) 135182}
  [\href{https://arxiv.org/abs/1908.02278}{{\ttfamily 1908.02278}}].

\bibitem{Dev:2020kgz}
A.~Dev, P.A.N.~Machado and P.~Mart\'\i{}nez-Mirav\'e, \emph{{Signatures of
  ultralight dark matter in neutrino oscillation experiments}},
  \href{https://doi.org/10.1007/JHEP01(2021)094}{\emph{JHEP} {\bfseries 01}
  (2021) 094} [\href{https://arxiv.org/abs/2007.03590}{{\ttfamily
  2007.03590}}].

\bibitem{Losada:2021bxx}
M.~Losada, Y.~Nir, G.~Perez and Y.~Shpilman, \emph{{Probing scalar dark matter
  oscillations with neutrino oscillations}},
  \href{https://doi.org/10.1007/JHEP04(2022)030}{\emph{JHEP} {\bfseries 04}
  (2022) 030} [\href{https://arxiv.org/abs/2107.10865}{{\ttfamily
  2107.10865}}].

\bibitem{Huang:2021kam}
G.-y.~Huang and N.~Nath, \emph{{Neutrino meets ultralight dark matter:
  0\ensuremath{\nu}\ensuremath{\beta}\ensuremath{\beta} decay and cosmology}},
  \href{https://doi.org/10.1088/1475-7516/2022/05/034}{\emph{JCAP} {\bfseries
  05} (2022) 034} [\href{https://arxiv.org/abs/2111.08732}{{\ttfamily
  2111.08732}}].

\bibitem{Chun:2021ief}
E.J.~Chun, \emph{{Neutrino Transition in Dark Matter}},
  \href{https://arxiv.org/abs/2112.05057}{{\ttfamily 2112.05057}}.

\bibitem{Dev:2022bae}
A.~Dev, G.~Krnjaic, P.~Machado and H.~Ramani, \emph{{Constraining feeble
  neutrino interactions with ultralight dark matter}},
  \href{https://doi.org/10.1103/PhysRevD.107.035006}{\emph{Phys. Rev. D}
  {\bfseries 107} (2023) 035006}
  [\href{https://arxiv.org/abs/2205.06821}{{\ttfamily 2205.06821}}].

\bibitem{Huang:2022wmz}
G.-y.~Huang, M.~Lindner, P.~Mart\'\i{}nez-Mirav\'e and M.~Sen,
  \emph{{Cosmology-friendly time-varying neutrino masses via the sterile
  neutrino portal}},
  \href{https://doi.org/10.1103/PhysRevD.106.033004}{\emph{Phys. Rev. D}
  {\bfseries 106} (2022) 033004}
  [\href{https://arxiv.org/abs/2205.08431}{{\ttfamily 2205.08431}}].

\bibitem{Losada:2022uvr}
M.~Losada, Y.~Nir, G.~Perez, I.~Savoray and Y.~Shpilman, \emph{{Parametric
  resonance in neutrino oscillations induced by ultra-light dark matter and
  implications for KamLAND and JUNO}},
  \href{https://doi.org/10.1007/JHEP03(2023)032}{\emph{JHEP} {\bfseries 03}
  (2023) 032} [\href{https://arxiv.org/abs/2205.09769}{{\ttfamily
  2205.09769}}].

\bibitem{Brzeminski:2022rkf}
D.~Brzeminski, S.~Das, A.~Hook and C.~Ristow, \emph{{Constraining Vector Dark
  Matter with neutrino experiments}},
  \href{https://doi.org/10.1007/JHEP08(2023)181}{\emph{JHEP} {\bfseries 08}
  (2023) 181} [\href{https://arxiv.org/abs/2212.05073}{{\ttfamily
  2212.05073}}].

\bibitem{Alonso-Alvarez:2023tii}
G.~Alonso-\'Alvarez, K.~Bleau and J.M.~Cline, \emph{{Distortion of neutrino
  oscillations by dark photon dark matter}},
  \href{https://doi.org/10.1103/PhysRevD.107.055045}{\emph{Phys. Rev. D}
  {\bfseries 107} (2023) 055045}
  [\href{https://arxiv.org/abs/2301.04152}{{\ttfamily 2301.04152}}].

\bibitem{Losada:2023zap}
M.~Losada, Y.~Nir, G.~Perez, I.~Savoray and Y.~Shpilman, \emph{{Time dependent
  CP-even and CP-odd signatures of scalar ultralight dark matter in neutrino
  oscillations}},
  \href{https://doi.org/10.1103/PhysRevD.108.055004}{\emph{Phys. Rev. D}
  {\bfseries 108} (2023) 055004}
  [\href{https://arxiv.org/abs/2302.00005}{{\ttfamily 2302.00005}}].

\bibitem{Davoudiasl:2023uiq}
H.~Davoudiasl and P.B.~Denton, \emph{{Sterile neutrino shape shifting caused by
  dark matter}}, \href{https://doi.org/10.1103/PhysRevD.108.035013}{\emph{Phys.
  Rev. D} {\bfseries 108} (2023) 035013}
  [\href{https://arxiv.org/abs/2301.09651}{{\ttfamily 2301.09651}}].

\bibitem{Gherghetta:2023myo}
T.~Gherghetta and A.~Shkerin, \emph{{Probing a local dark matter halo with
  neutrino oscillations}},
  \href{https://doi.org/10.1103/PhysRevD.108.095009}{\emph{Phys. Rev. D}
  {\bfseries 108} (2023) 095009}
  [\href{https://arxiv.org/abs/2305.06441}{{\ttfamily 2305.06441}}].

\bibitem{ChoeJo:2023ffp}
Y.~ChoeJo, Y.~Kim and H.-S.~Lee, \emph{{Dirac-Majorana neutrino type
  oscillation induced by a wave dark matter}},
  \href{https://doi.org/10.1103/PhysRevD.108.095028}{\emph{Phys. Rev. D}
  {\bfseries 108} (2023) 095028}
  [\href{https://arxiv.org/abs/2305.16900}{{\ttfamily 2305.16900}}].

\bibitem{Chen:2023vkq}
Y.~Chen, X.~Xue and V.~Cardoso, \emph{{Black Holes as Neutrino Factories}},
  \href{https://arxiv.org/abs/2308.00741}{{\ttfamily 2308.00741}}.

\bibitem{Hui:2021tkt}
L.~Hui, \emph{{Wave Dark Matter}},
  \href{https://doi.org/10.1146/annurev-astro-120920-010024}{\emph{Ann. Rev.
  Astron. Astrophys.} {\bfseries 59} (2021) 247}
  [\href{https://arxiv.org/abs/2101.11735}{{\ttfamily 2101.11735}}].

\bibitem{Preskill:1982cy}
J.~Preskill, M.B.~Wise and F.~Wilczek, \emph{{Cosmology of the Invisible
  Axion}}, \href{https://doi.org/10.1016/0370-2693(83)90637-8}{\emph{Phys.
  Lett. B} {\bfseries 120} (1983) 127}.

\bibitem{Abbott:1982af}
L.F.~Abbott and P.~Sikivie, \emph{{A Cosmological Bound on the Invisible
  Axion}}, \href{https://doi.org/10.1016/0370-2693(83)90638-X}{\emph{Phys.
  Lett. B} {\bfseries 120} (1983) 133}.

\bibitem{Dine:1982ah}
M.~Dine and W.~Fischler, \emph{{The Not So Harmless Axion}},
  \href{https://doi.org/10.1016/0370-2693(83)90639-1}{\emph{Phys. Lett. B}
  {\bfseries 120} (1983) 137}.

\bibitem{ParticleDataGroup:2022pth}
{\scshape Particle Data Group} collaboration, \emph{{Review of Particle
  Physics}}, \href{https://doi.org/10.1093/ptep/ptac097}{\emph{PTEP} {\bfseries
  2022} (2022) 083C01}.

\bibitem{Bi:2003yr}
X.-J.~Bi, P.-h.~Gu, X.-l.~Wang and X.-m.~Zhang, \emph{{Thermal leptogenesis in
  a model with mass varying neutrinos}},
  \href{https://doi.org/10.1103/PhysRevD.69.113007}{\emph{Phys. Rev. D}
  {\bfseries 69} (2004) 113007}
  [\href{https://arxiv.org/abs/hep-ph/0311022}{{\ttfamily hep-ph/0311022}}].

\bibitem{Gu:2004xx}
P.-h.~Gu and X.-j.~Bi, \emph{{Thermal leptogenesis with triplet Higgs boson and
  mass varying neutrinos}},
  \href{https://doi.org/10.1103/PhysRevD.70.063511}{\emph{Phys. Rev. D}
  {\bfseries 70} (2004) 063511}
  [\href{https://arxiv.org/abs/hep-ph/0405092}{{\ttfamily hep-ph/0405092}}].

\bibitem{Hati:2015hvq}
C.~Hati and U.~Sarkar, \emph{{Neutrino dark energy and leptogenesis with TeV
  scale triplets}},
  \href{https://doi.org/10.1140/epjc/s10052-016-4089-6}{\emph{Eur. Phys. J. C}
  {\bfseries 76} (2016) 236}
  [\href{https://arxiv.org/abs/1511.02874}{{\ttfamily 1511.02874}}].

\bibitem{Bae:2016zym}
K.J.~Bae, H.~Baer, K.~Hamaguchi and K.~Nakayama, \emph{{Affleck-Dine
  Leptogenesis with Varying Peccei-Quinn Scale}},
  \href{https://doi.org/10.1007/JHEP02(2017)017}{\emph{JHEP} {\bfseries 02}
  (2017) 017} [\href{https://arxiv.org/abs/1612.02511}{{\ttfamily
  1612.02511}}].

\bibitem{Turner:1983he}
M.S.~Turner, \emph{{Coherent Scalar Field Oscillations in an Expanding
  Universe}}, \href{https://doi.org/10.1103/PhysRevD.28.1243}{\emph{Phys. Rev.
  D} {\bfseries 28} (1983) 1243}.

\bibitem{Doi:1985dx}
M.~Doi, T.~Kotani and E.~Takasugi, \emph{{Double beta Decay and Majorana
  Neutrino}}, \href{https://doi.org/10.1143/PTPS.83.1}{\emph{Prog. Theor. Phys.
  Suppl.} {\bfseries 83} (1985) 1}.

\bibitem{GERDA:2022ffe}
{\scshape GERDA} collaboration, \emph{{Search for exotic physics in
  double-\ensuremath{\beta} decays with GERDA Phase~II}},
  \href{https://doi.org/10.1088/1475-7516/2022/12/012}{\emph{JCAP} {\bfseries
  12} (2022) 012} [\href{https://arxiv.org/abs/2209.01671}{{\ttfamily
  2209.01671}}].

\bibitem{DeSimone:2006nrs}
A.~De~Simone and A.~Riotto, \emph{{On the impact of flavour oscillations in
  leptogenesis}},
  \href{https://doi.org/10.1088/1475-7516/2007/02/005}{\emph{JCAP} {\bfseries
  02} (2007) 005} [\href{https://arxiv.org/abs/hep-ph/0611357}{{\ttfamily
  hep-ph/0611357}}].

\bibitem{Blanchet:2006ch}
S.~Blanchet, P.~Di~Bari and G.G.~Raffelt, \emph{{Quantum Zeno effect and the
  impact of flavor in leptogenesis}},
  \href{https://doi.org/10.1088/1475-7516/2007/03/012}{\emph{JCAP} {\bfseries
  03} (2007) 012} [\href{https://arxiv.org/abs/hep-ph/0611337}{{\ttfamily
  hep-ph/0611337}}].

\bibitem{Blanchet:2011xq}
S.~Blanchet, P.~Di~Bari, D.A.~Jones and L.~Marzola, \emph{{Leptogenesis with
  heavy neutrino flavours: from density matrix to Boltzmann equations}},
  \href{https://doi.org/10.1088/1475-7516/2013/01/041}{\emph{JCAP} {\bfseries
  01} (2013) 041} [\href{https://arxiv.org/abs/1112.4528}{{\ttfamily
  1112.4528}}].

\bibitem{Moffat:2018wke}
K.~Moffat, S.~Pascoli, S.T.~Petcov, H.~Schulz and J.~Turner,
  \emph{{Three-flavored nonresonant leptogenesis at intermediate scales}},
  \href{https://doi.org/10.1103/PhysRevD.98.015036}{\emph{Phys. Rev. D}
  {\bfseries 98} (2018) 015036}
  [\href{https://arxiv.org/abs/1804.05066}{{\ttfamily 1804.05066}}].

\bibitem{Granelli:2021fyc}
A.~Granelli, K.~Moffat and S.T.~Petcov, \emph{{Aspects of high scale
  leptogenesis with low-energy leptonic CP violation}},
  \href{https://doi.org/10.1007/JHEP11(2021)149}{\emph{JHEP} {\bfseries 11}
  (2021) 149} [\href{https://arxiv.org/abs/2107.02079}{{\ttfamily
  2107.02079}}].

\bibitem{Granelli:2023egb}
A.~Granelli, K.~Hamaguchi, N.~Nagata, M.E.~Ramirez-Quezada and J.~Wada,
  \emph{{Thermal leptogenesis in the minimal gauged $ \textrm{U}{(1)}_{L_{\mu
  }-{L}_{\tau }} $ model}},
  \href{https://doi.org/10.1007/JHEP09(2023)079}{\emph{JHEP} {\bfseries 09}
  (2023) 079} [\href{https://arxiv.org/abs/2305.18100}{{\ttfamily
  2305.18100}}].

\bibitem{Buchmuller:2002rq}
W.~Buchmuller, P.~Di~Bari and M.~Plumacher, \emph{{Cosmic microwave background,
  matter - antimatter asymmetry and neutrino masses}},
  \href{https://doi.org/10.1016/S0550-3213(02)00737-X}{\emph{Nucl. Phys. B}
  {\bfseries 643} (2002) 367}
  [\href{https://arxiv.org/abs/hep-ph/0205349}{{\ttfamily hep-ph/0205349}}].

\bibitem{Buchmuller:2004nz}
W.~Buchmuller, P.~Di~Bari and M.~Plumacher, \emph{{Leptogenesis for
  pedestrians}}, \href{https://doi.org/10.1016/j.aop.2004.02.003}{\emph{Annals
  Phys.} {\bfseries 315} (2005) 305}
  [\href{https://arxiv.org/abs/hep-ph/0401240}{{\ttfamily hep-ph/0401240}}].

\bibitem{Davidson:2002qv}
S.~Davidson and A.~Ibarra, \emph{{A Lower bound on the right-handed neutrino
  mass from leptogenesis}},
  \href{https://doi.org/10.1016/S0370-2693(02)01735-5}{\emph{Phys. Lett. B}
  {\bfseries 535} (2002) 25}
  [\href{https://arxiv.org/abs/hep-ph/0202239}{{\ttfamily hep-ph/0202239}}].

\bibitem{Pilaftsis:1998pd}
A.~Pilaftsis, \emph{{Heavy Majorana neutrinos and baryogenesis}},
  \href{https://doi.org/10.1142/S0217751X99000932}{\emph{Int. J. Mod. Phys. A}
  {\bfseries 14} (1999) 1811}
  [\href{https://arxiv.org/abs/hep-ph/9812256}{{\ttfamily hep-ph/9812256}}].

\bibitem{Pilaftsis:2003gt}
A.~Pilaftsis and T.E.J.~Underwood, \emph{{Resonant leptogenesis}},
  \href{https://doi.org/10.1016/j.nuclphysb.2004.05.029}{\emph{Nucl. Phys. B}
  {\bfseries 692} (2004) 303}
  [\href{https://arxiv.org/abs/hep-ph/0309342}{{\ttfamily hep-ph/0309342}}].

\bibitem{Pontecorvo:1957cp}
B.~Pontecorvo, \emph{{Mesonium and anti-mesonium}}, {\emph{Sov. Phys. JETP}
  {\bfseries 6} (1957) 429}.

\bibitem{Maki:1962mu}
Z.~Maki, M.~Nakagawa and S.~Sakata, \emph{{Remarks on the unified model of
  elementary particles}}, \href{https://doi.org/10.1143/PTP.28.870}{\emph{Prog.
  Theor. Phys.} {\bfseries 28} (1962) 870}.

\bibitem{Planck:2018vyg}
{\scshape Planck} collaboration, \emph{{Planck 2018 results. VI. Cosmological
  parameters}},
  \href{https://doi.org/10.1051/0004-6361/201833910}{\emph{Astron. Astrophys.}
  {\bfseries 641} (2020) A6}
  [\href{https://arxiv.org/abs/1807.06209}{{\ttfamily 1807.06209}}].

\bibitem{SimonsObservatory:2018koc}
{\scshape Simons Observatory} collaboration, \emph{{The Simons Observatory:
  Science goals and forecasts}},
  \href{https://doi.org/10.1088/1475-7516/2019/02/056}{\emph{JCAP} {\bfseries
  02} (2019) 056} [\href{https://arxiv.org/abs/1808.07445}{{\ttfamily
  1808.07445}}].

\bibitem{Abazajian:2019eic}
K.~Abazajian et~al., \emph{{CMB-S4 Science Case, Reference Design, and Project
  Plan}},  \href{https://arxiv.org/abs/1907.04473}{{\ttfamily 1907.04473}}.

\end{thebibliography}\endgroup

\end{document}